\documentclass[nouppercase]{ifmbe}

\usepackage[cmex10]{amsmath}
\usepackage{amssymb}
\usepackage{lineno}

\title{Bayesian Modelling of Skull Conductivity Uncertainties in EEG Source Imaging}

\affiliation{Institute for Biomagnetism and Biosignalanalysis, University of M\"unster, M\"unster, Germany }{FIRSTAFF}
\affiliation{Institute for Computational and Applied Mathematics, University of M\"unster, M\"unster, Germany }{SECONDAFF}
\affiliation{Centre for Medical Image Computing, University College London, London, UK }{THA}
\affiliation{Department of Mathematics, University of Auckland, Auckland, New Zealand }{FOA}

\author{V. Rimpil\"ainen}{FIRSTAFF}
\author{A. Koulouri}{SECONDAFF}
\author{F. Lucka}{THA}
\author{J. P. Kaipio}{FOA}
\author{C. H. Wolters}{FIRSTAFF}
 
\begin{document}

\maketitle

\begin{abstract}
Knowing the correct skull conductivity is crucial for the accuracy of EEG source imaging, but unfortunately, its true value, which is inter- and intra-individually varying, is difficult to determine. In this paper, we propose a statistical method based on the Bayesian approximation error approach to compensate for source imaging errors related to erronous skull conductivity. We demonstrate the potential of the approach by simulating EEG data of focal source activity and using the dipole scan algorithm and a sparsity promoting prior to reconstruct the underlying sources. The results suggest that the greatest improvements with the proposed method can be achieved when the focal sources are close to the skull.
\end{abstract}



\section{Introduction}

Source reconstruction from electroencephalography (EEG) data is an ill-posed inverse problem, and the solution depends strongly on the accuracy of the discretized head model \cite{Ellenrieder2006,Vanrumste2000,vor14}. The essential head features are the geometry and the electrical properties of the tissues of which the geometry can be extracted, to some extent, using auxiliary imaging tools \cite{Slomka2009}.

However, the accurate determination of the electrical conductivities of the head tissues is an open question. Especially the accurate modelling of the skull is essential for source reconstruction. There are only few techniques available and they can be used to calibrate only few tissue conductivity parameters either by using well defined somatosensory evoked potentials / fields in combination with EEG \cite{lew09b}, combined EEG/MEG \cite{wol10,ayd14} or electrical impedance tomography (EIT) -based techniques \cite{gon03a}. Since all these methods rely on auxiliary measurements, it would be highly beneficial to find methods that are purely computational for this purpose.

In this paper, we propose to use the Bayesian approximation error (BAE) approach to compensate for the imaging artefacts caused by the approximately modelled skull conductivity. The main idea of BAE is to use an approximate model (e.g. because the accurate model is unknown or computationally unfeasible) and to take statistically into account the induced modelling errors \cite{kaipio2013}. In our case, we first choose an observation model with a standard skull conductivity value and then estimate statistics of the errors between this approximate model and the set of accurate models that is constructed based on the postulated distribution of possible skull conductivity values. This results in an additive error term that is marginalized using a Gaussian approximation. Previously, BAE has been successfully used in EIT \cite{Nissinen2009}, optical tomography \cite{Arridge2006} and recently also in EEG source imaging \cite{kou16} for geometry related modelling errors.


We demonstrate here by using a specific BAE model, namely the enhanced error model (EEM), that the imaging errors caused by the erroneous skull conductivity can be alleviated. The work was carried out by using state-of-the-art finite element (FE) -based head models, skull conductivity values based on the known literature and appropriate sparsity priors to recover focal source configurations.


\section{Theory}

\subsection{Bayesian framework with linear forward model}

The computational domain is denoted with $\Omega$ and its material properties with $\sigma(x)$ where $x \in \Omega$. For numerical implementations, the domain is discretized and the observation model is written as 
\begin{equation}\label{eq:obm2}
v = A(\sigma) d + e,
\end{equation}
where $v \in \mathbb{R}^m$ are the measurements, $m$ is the number of measurements, $A(\sigma) \in \mathbb{R}^{m \times 3n}$ is the lead field matrix that depends on electric conductivity $\sigma$, $d\in \mathbb{R}^{3n}$ is the distributed dipole source configuration and $e \sim \mathcal{N}(e_*,\Gamma_e)$ is the measurement noise. 
Note that the model $A(\sigma)$ assumes that the accurate values of electric conductivities are known (which in practice without additional effort is almost never the case).

In the Bayesian framework, the inverse
solution is the posterior density of the Bayes’ formula
\begin{equation}\label{post}
\pi(d|v) \propto \pi(v|d)\pi(d),
\end{equation}
where $\pi(v|d)$ is the likelihood and $\pi(d)$ the prior. 

The likelihood model can be written as
\begin{equation}\label{lik1}
\pi(v|d) \propto \exp\Big(-\frac{1}{2}(v-Ad-e_{*})^\mathrm{T}\Gamma_e^{-1}(v-Ad-e_{*})\Big).
\end{equation}

\subsection{Enhanced error model}

In BAE, we replace the accurate lead field matrix, $A$, with an approximate lead field,
$A_{0}$, in which we employ standard fixed values for the electric conductivity, $\sigma_0$.
We can write
\begin{equation}\label{eq:AMEA}
v= A_{0}d+\varepsilon+e
\end{equation}
where $\varepsilon = Ad-A_0d$ is the induced {\em approximation error}, $\varepsilon \in \mathbb{R}^m$. 

In the enhanced error model (EEM), we further approximate that $\varepsilon \sim \mathcal{N}(\varepsilon_*,\Gamma_\varepsilon)$ and $d$ are independent, i.e., $\varepsilon$ is considered as another random additive error term. Even though, in practice, $\varepsilon$ and $d$ are usually correlated, this approximation often leads to very similar inverse solutions \cite{kaipio2013}. In our case, since we assume that the true source activity $d$ is focal (sparse), the cross-covariances with $\varepsilon$ will be negligible. 

Based on these approximations, we formulate the probability distribution of the likelihood as 
\begin{eqnarray}\label{lik2}
&&\pi(v | d) \propto 
\exp
\Big(-\frac{1}{2}(v-A_{0}d-\varepsilon_{*}-e_{*})^\mathrm{T} \nonumber \\
&&\quad \qquad \qquad (\Gamma_{\varepsilon}+\Gamma_e)^{-1}(v-A_{0}d-\varepsilon_{*}-e_{*})\Big).
\end{eqnarray}

\subsection{Prior model}
\label{prior}
In this paper, we consider sparse focal source models and we employ an $L_{1,2}$ norm induced 
prior model (group lasso)
\begin{equation}
\pi(d)\propto\exp{\left({-\frac{\alpha}{2}\sum_{i=1}^{n}w_i\|d_{i}\|_2}\right)}
\label{l12}
\end{equation}
where $d_{i}=(d_{ix},d_{iy},d_{iz})$ and $\|d_{i}\|^{1}_2=\sqrt{d_{ix}^2+d_{iy}^2+d_{iz}^2}$ is the strength of the source at location $i$, and $\alpha$ is a scaling constant 
that fixes the distribution of the source strengths. The depth weighting factors $w_i$ are used to reduce the bias of the prior to
favor superficial source distributions \cite{Haufe2008,kou16}.

\subsection{Maximum a posteriori estimate}
%
In this paper, we compute maximum a posteriori (MAP) estimates of the posterior.
Based on the likelihood model (\ref{lik1}), the MAP estimate of the source configuration is 
\begin{equation}\label{eq:Functional1}
\hat{d}_{\mathrm{ACC}}=\min_d  \|L_{e}(v-Ad-e_{*})\|_2^2 +
\alpha\sum_{i=1}^{n}w_i\|d_{i}\|_2 ,
\end{equation}
where $L_e$ comes from the Cholesky factorization $\Gamma_{e}^{-1}=L_e^T L_e$. The subscript "ACC" refers to the assumption that the electrical conductivities are accurately known.

If we, however, compute the solution using the fixed standard electrical conductivity values, $\sigma_0$, we get
\begin{equation}\label{eq:Functional2}
\hat{d}_{\mathrm{STAN}}=\min_d\|L_{e}(v-A_0d-e_{*})\|_2^2 +
\alpha\sum_{i=1}^{n}w_i\|d_{i}\|_2.
\end{equation}

From the likelihood of the EEM (\ref{lik2}), the source configuration can be estimated as
\begin{equation}\label{eq:Functional3}
\hat{d}_{\mathrm{EEM}}=\min_d\|L_{\varepsilon+e}(v-A_0d-\varepsilon_*-e_{*})\|_2^2 +
\alpha\sum_{i=1}^{n}w_i\|d_{i}\|_2
\end{equation}
where $(\Gamma_{\varepsilon}+\Gamma_{e})^{-1}=L_{\varepsilon+e}^T L_{\varepsilon+e}$. Note that the EEM result is computed using the same lead field matrix as in Equation (\ref{eq:Functional2}).

The resulting non-linear convex minimization problem is solved using the truncated Newton
interior point method \cite{Boyd2004}.


\section{Materials and methods}

\subsection{Head models}
\label{hm}
The geometry of the head was constructed based on T1- and T2-weighted magnetic resonance images of a healthy subject measured with a 3 T MR scanner. The scalp, eyes, skull compacta, skull spongiosa, cerebro spinal fluid (CSF), gray matter (GM) and white matter (WM) were segmented, for more details see \cite{luc12,ayd14}. In the simulation set-up, 74 measurement electrodes were attached to the sculp.

The electric conductivities (in S/m) of the different tissues were 0.43 for the scalp \cite{dan11}, 0.505 for the eyes \cite{ram06}, 1.79 for the CSF \cite{bau97}, 0.14 for the WM \cite{ram06} and 0.33 for the GM \cite{ram06}. The skull conductivities of the different head models were the following:

First, we created 200 head models with skull conductivity drawn from a Gaussian distribution with mean 0.01855 and standard deviation 0.007225. This distribution was set in such a way that the two standard deviation lower and upper values were 0.0041 \cite{hom95,ayd14,fuc98,buc97} and 0.033 \cite{hoe03,ayd14}, respectively. We refer to these head models as {\it sample} head model. 

We also created a \textit{standard} head model with the mean skull conductivity 0.01855. The skull spongiosa conductivities in all the models were selected based on the spongiosa:compacta conductivity ratio 3.6:1 \cite{ayd14}.

The head geometry was discretized using tetrahedral FEs. For the forward simulations, a source space that covered the GM was constructed with 30,105 nodes on a regular grid with grid size 2 mm. For the inverse computations, two source spaces that covered the GM were used: the first one consisted of 10,782 source locations on a regular grid with grid size 3 mm and the second 2,249 source locations with grid size 5 mm. The forward grid was chosen in such a way that it did not contain the same coordinate points as the inverse grids. The lead field matrices used in the simulations were computed using standard piecewise linear FE basis functions with the Saint Venant source modelling approach \cite{buc97,wol07}.

\subsection{Computation of the approximation error statistics}
\label{ces} 
The approximation error statistics was created by first choosing randomly one of the sample head models, evaluating the model with randomly chosen source configuration, and finally calculating the approximation error by evaluating the standard model with the same source configuration, 
\begin{equation}
\varepsilon^{(j)} = A(\sigma^{(j)}){d}^{(j)} -
A_{0}d^{(j)},
\end{equation}
where $A(\sigma^{(j)})$ is one of the sample models, ${d}^{(j)}$ random source configuration and $A_{0}$ is the standard model.

The procedure was repeated $J=200,000$ times, and these simulated error samples were used to calculate the sample mean, $\varepsilon_*$, and the sample covariance, $\Gamma_{\varepsilon}$, of the approximation error.


\section{Results and discussion}

\subsection{Dipole scan with noiseless data}

To demonstrate the potential of BAE, we first carried out reconstructions from noiseless data with the dipole scan algorithm. The noiseless data was computed using the {\it accurate} model that had skull conductivity 0.0041 S/m \cite{hom95,ayd14,fuc98,buc97}. This value was chosen to showcase a rather large difference compared to the skull conductivity in the standard model, 0.01855 S/m. Single sources in GM with orientation that was normal to the surface of the cortex were used. For the dipole scan, the 3 mm regular grid was used and two results were computed: one with the standard model and another with the EEM. 

\begin{figure}[ht]
      \centering
          \includegraphics[width=0.99\columnwidth]{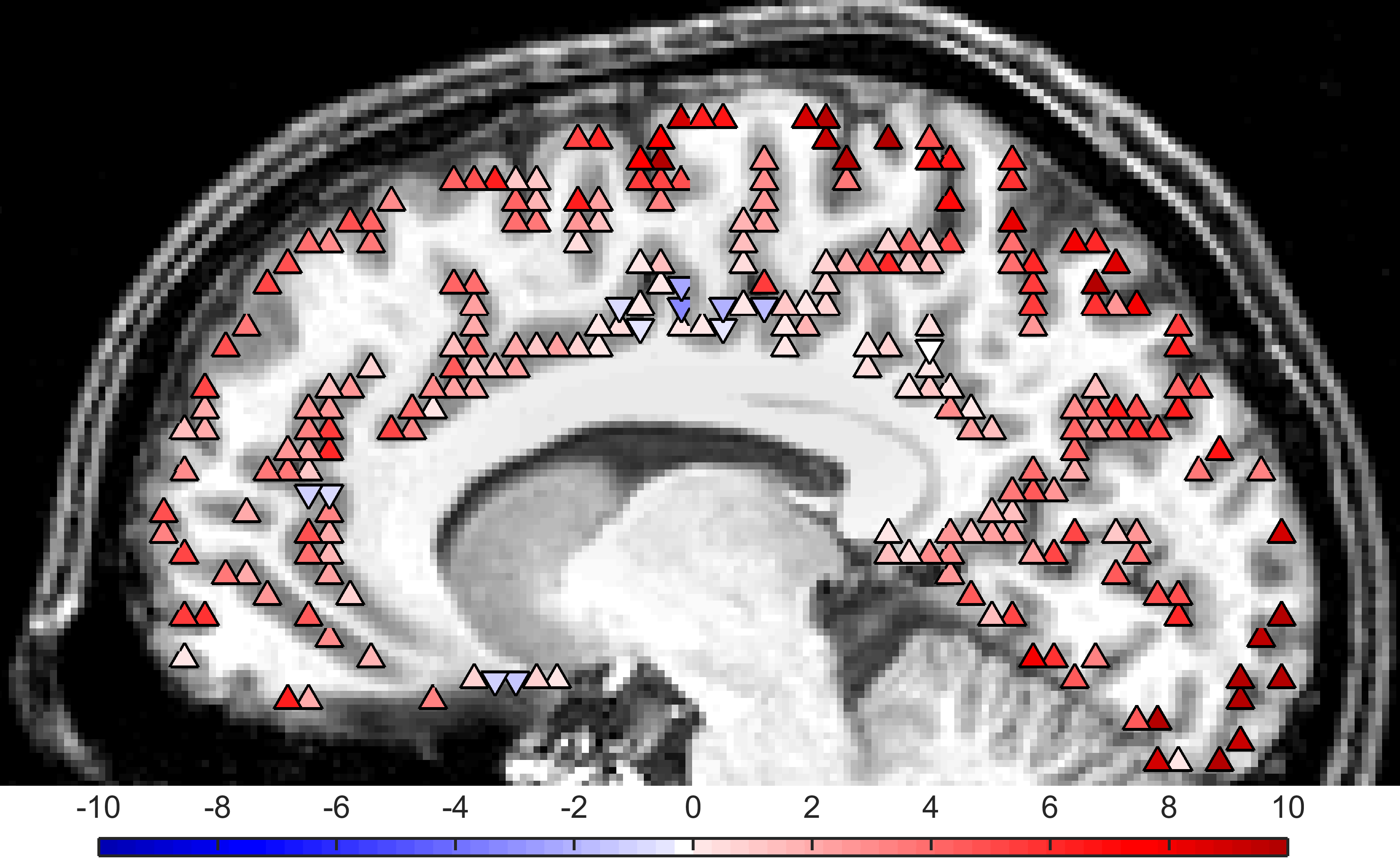}
      \caption{The red $\Delta$ signs show how many milli meters the proposed EEM improves the localization accuracy of the sources when compared to the standard model. The blue $\nabla$ signs show the locations where the EEM has less accuracy than the standard model.}
      \label{figure1}
\end{figure}

From the results, we calculated the localization error (in milli meters) between the actual and the reconstructed source for both models, $X_{STAN}$ and $X_{EEM}$, respectively. Figure \ref{figure1} shows the differences $\Delta = X_{STAN} - X_{EEM}$ which are positive (red $\Delta$ signs) if EEM has improved the source localization, close to zero (white $\Delta$ signs) if both models give the same localization error, and negative (blue $\nabla$ signs) if EEM performs worse than the standard model.

It can be seen that the largest improvements occur for sources that are close to the skull. These are also the sources that usually have the largest localization errors due to erronous skull conductivity. The deeper in the brain the sources are the less improvement can be seen, and for some deep locations the EEM model actually performs worse than the standard model.

\subsection{$L_{1,2}$ prior with noisy data}

\begin{figure*}[htb]
      \centering
          \includegraphics[width=2\columnwidth]{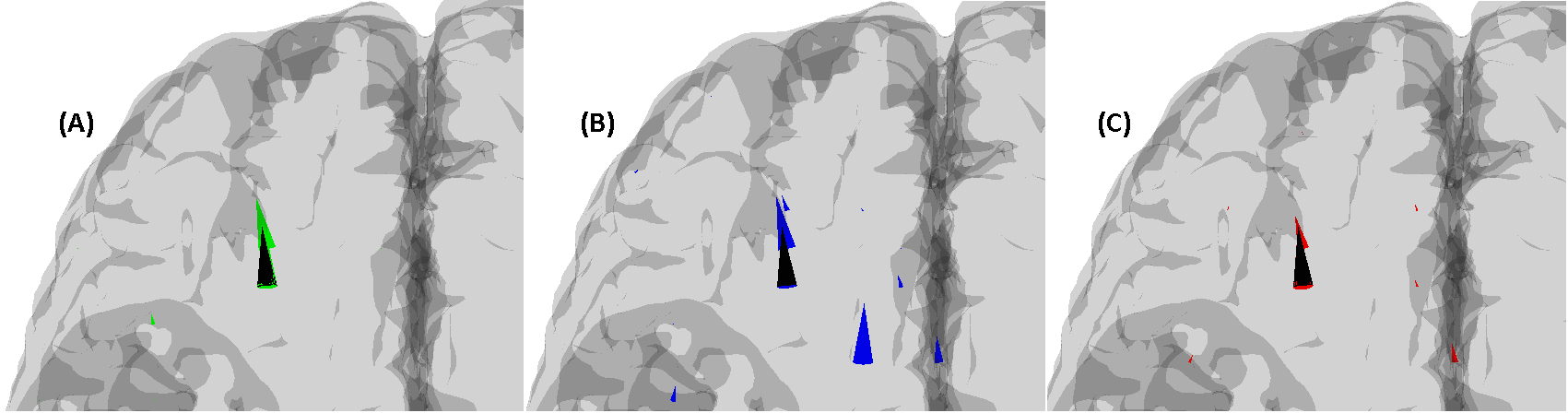}
      \caption{(A) The location of the source is shown with a black cone and the reconstruction result when using the accurate model with green cones. (B) The reconstruction result using the standard model (with wrong skull conductivity) is shown with blue cones. (C) The EEM reconstruction, red cones, gives a better estimate for the sources than the standard model.}
      \label{figure2}
\end{figure*}

To demonstrate the feasibility with a sparsity prior, we present a result computed from noisy boundary data (SNR = 40 dB) by using the $L_{1,2}$ prior described in Section II-\ref{prior}. Here, the results were computed by evaluating Equations (\ref{eq:Functional1})-(\ref{eq:Functional3}), respectively. The computations were carried out in the regular 5 mm grid. The first result in Figure \ref{figure2} shows the correct location of the source (black cone) and the reconstruction when the accurate skull conductivity, 0.0041 S/m, is known (green cones). The second result shows the reconstruction when the standard model is used (blue cones) with skull conductivity 0.01855, and the third result the reconstruction with the EEM (red cones). As expected, the best results is achieved when the accurate skull conductivity is known. However, if the accurate conductivity is not know, then EEM gives a better estimate than the standard model.

\section{Conclusion and future work}

We have demonstrated that with the help of EEM it is possible to reduce source localization errors caused by erroneous skull conductivity. The EEM showed the greatest improvement for sources close to the skull. In the future, the statistical modelling of conductivity errors can be combined with various geometry related approximations.

\section*{Conflict of interest}

The authors declare that they have no conflict of interest.

\bibliography{rimpilainen}

\begin{table}[h]
\footnotesize
        \begin{tabular}{ll}
        &Corresponding author: Ville Rimpil\"ainen \\
        &Institute: University of M\"unster\\
        &Street: Malmedyweg 15\\
        &City: M\"unster\\
        &Country:  Germany\\
        &Email: vrimpila@gmail.com \\
        \end{tabular}
\end{table}

\end{document}